\documentclass{article}

\PassOptionsToPackage{numbers}{natbib}

\usepackage[preprint]{neurips_2025}

\usepackage[utf8]{inputenc} 
\usepackage[T1]{fontenc}    
\usepackage{hyperref}       
\usepackage{url}            
\usepackage{booktabs}       
\usepackage{amsfonts}       
\usepackage{nicefrac}       
\usepackage{microtype}      
\usepackage{xcolor}         
\usepackage{graphicx}
\usepackage{subfigure}
\usepackage{booktabs} 
\usepackage{multirow}
\usepackage{bbding}
\usepackage{enumitem}
\usepackage{cleveref}
\usepackage{wrapfig}

\usepackage{xspace}

\newcommand{\ie}{\hbox{\emph{i.e.}}\xspace}

\newcommand{\F}{Figure}
\newcommand{\myhigh}[1]{\colorbox{gray!20}{\textbf{#1}}}

\title{Is Your Benchmark Still Useful?\\Dynamic Benchmarking for Code Language Models}

\author{%
  Batu Guan\\
  The Chinese University of Hong Kong \\
  \texttt{btguan@cse.cuhk.edu.hk} \\
  \And
    Xiao Wu\\
  Huazhong University of Science and Technology \\
  \texttt{xiaowu@hust.edu.cn} \\
  \And
  Yuanyuan Yuan \\
  ETH Zurich \\
  \texttt{yuanyuan.yuan@inf.ethz.ch} \\
  \And
  Shaohua Li \\
  The Chinese University of Hong Kong \\
  \texttt{shaohuali@cse.cuhk.edu.hk} \\
}

\begin{document}

\maketitle

\begin{abstract}
In this paper, we tackle a critical challenge in model evaluation: how to keep code benchmarks useful when models might have already seen them during training.
We introduce a novel solution, dynamic benchmarking framework, to address this challenge.
Given a code understanding or reasoning benchmark, our framework dynamically transforms each input, \ie, programs, with various semantic-preserving mutations to build a syntactically new while semantically identical benchmark.
We evaluated 10 popular language models on our dynamic benchmarks.
Our evaluation reveals several interesting or surprising findings:
(1) all models perform significantly worse than before,
(2) the ranking between some models shifts dramatically,
and (3) 
dynamic benchmarks can resist against the data contamination problem.

\end{abstract}

\section{Introduction}
\label{introduction}

During the past period, the emergence of advanced Large Language Models (LLMs) such as the GPT~\cite{hurst2024gpt}, Llama~\cite{dubey2024llama}, and DeepSeek~\cite{deepseekai2024deepseekv3technicalreport} series have revolutionized the field of software development through their exceptional performance in various code-related tasks, including but not limited to code generation~\cite{jiang2024surveylargelanguagemodels}, debugging~\cite{levin2024chatdbgaipowereddebuggingassistant}, and optimization~\cite{ishibashi2024selforganizedagentsllmmultiagent}. Rapid integration of LLM-driven development tools and plugins into mainstream programming environments, exemplified by GitHub Copilot~\cite{friedman2021introducing} and Cursor~\cite{cursor2023}, not only validates the practical utility of these models, but also significantly enhances the productivity of developers in the software industry. This widespread adoption ultimately underscores the transformative potential of LLMs in code understanding and reasoning, where LLMs are expected to comprehend the underlying semantics in the input code~\cite{liu2024codemindframeworkchallengelarge}.

During the rapid advancement of LLMs, the community has developed various benchmarks to evaluate the code reasoning capabilities of the model. For instance, CRUXEval~\cite{gu2024cruxeval} specifically targets code execution proficiency, while Code Lingua~\cite{pan2024lost} and TransCoder~\cite{lachaux2020unsupervisedtranslationprogramminglanguages, yang2024exploringunleashingpowerlarge} focus on evaluating code translation skills. The successful completion of these tasks requires LLMs to demonstrate accurate comprehension of code semantics; hence, these benchmarks serve as effective indicators of the LLMs' code reasoning capabilities to a certain extent.


There are a few common practices to collect code for a benchmark. 
Some benchmarks automatically crawl code from public databases such as GitHub~\cite{hui2024qwen25codertechnicalreport,lozhkov2024starcoder2stackv2}, while others take advantage of advanced LLMs, such as ChatGPT, to generate new code~\cite{gu2024cruxeval}. In some cases, human experts are involved in the manual writing of the code. 
Regardless of how the code is collected, all these open-source benchmarks will eventually be available to the public for fair model evaluation.
Modern LLMs are trained on large amounts of data, both public and private. 
For instance, Qwen2.5-Coder's report states that they collect publicly available repositories on GitHub for training and remove key datasets using a 10-gram overlap method~\cite{hui2024qwen25codertechnicalreport}. However, removing key datasets does not guarantee that all data used for evaluation is eliminated, and there is still a risk of data contamination.
This raises a crucial concern: 

\emph{Are these benchmarks still useful for evaluating models if they are unavoidably included in the training data?}


\begin{figure*}[t]
\begin{center}
\begin{tabular}{cc}
\subfigure[Original Version]{
    \includegraphics[width=0.45\textwidth]{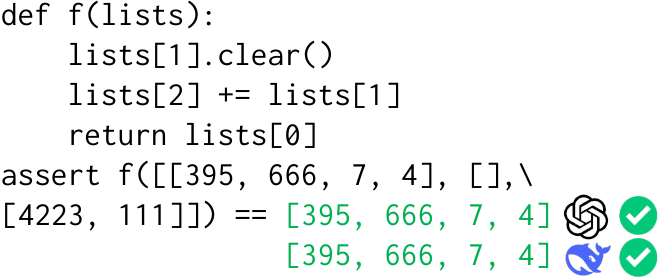}
    \label{fig:fig1_1} 
} &
\subfigure[Mutated Version]{
    \includegraphics[width=0.45\textwidth]{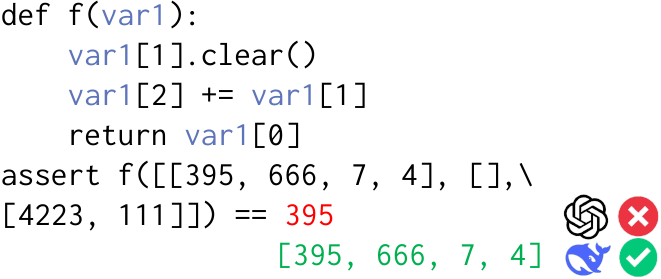}
    \label{fig:fig1_2} 
} \\
\end{tabular}
\caption{Answers produced by GPT-4o mini \raisebox{-0.5pt}{\includegraphics[width=0.4cm]{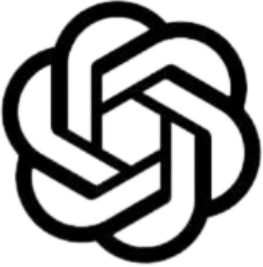}} and DeepSeek V3 \raisebox{-0.1pt}{\includegraphics[width=0.5cm]{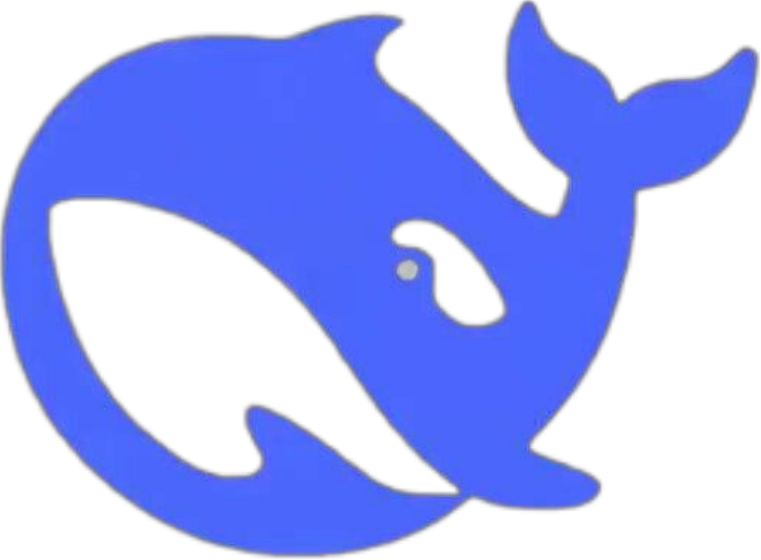}}. The modified variables are highlighted in blue for identification.}
\label{fig:fig1} 
\end{center}
\vspace{-10pt}
\end{figure*}

Take \F~\ref{fig:fig1} as an example. On the left is a test case from the CRUXEval benchmark~\cite{gu2024cruxeval}. The test case consists of a function \texttt{f} together with the input \texttt{([[395, 666, 7, 4], [], [4223, 111]])} --- a list of lists --- encoded in an \texttt{assert}. Models are tasked to predict the function's output.
The function \texttt{f} executes a sequence of operations: first erasing the value in \texttt{lists[1]}, then adding this value to \texttt{lists[2]}, and finally outputting the value of \texttt{lists[0]}. GPT-4o mini successfully emits the correct output in this scenario. However, as shown in \F~\ref{fig:fig1_2}, after we change the variable name into \texttt{var1}, GPT-4o mini produces an incorrect result with even the wrong type (an integer instead of a list of integers). 
This behavioral shift suggests that in \F~\ref{fig:fig1_1}, the descriptive variable name \texttt{lists} may have 
assisted the model's comprehension, {potentially due to the contamination of variable name \texttt{lists} in the training data.}
In \F~\ref{fig:fig1_2}, however, the absence of such hints, combined with the common pattern of \texttt{var1[0]} typically representing integer indices, likely led the model to erroneously infer an integer return type. 
DeepSeek V3, on the other hand, successfully handles both cases, {indicating its reasoning beyond variable names.} Nevertheless, the original benchmark fails to differentiate these two models' {reasoning capabilities} with this test case.



When the code in \F~\ref{fig:fig1_1} appears in the training set of a model, such code cannot be effectively used to measure the code understanding and reasoning capability of the model. 
Unfortunately, publicly accessible benchmarks are persistently plagued by the issue of training data contamination, which fundamentally compromises the integrity and reliability of evaluation results~\cite{aiyappa2023can}. This phenomenon is particularly problematic given the remarkable memorization capacity of LLMs~\cite{hartmann2023sokmemorizationgeneralpurposelarge}, where models may have already been exposed to substantial portions of the evaluation data\footnote{{Or code corpus of similar syntactic patterns.}}, which are open source online, during their training phase. Such exposure enables models to potentially reproduce or closely approximate memorized patterns rather than demonstrating genuine reasoning or generalization capabilities. 
Consequently, this leads to inflating performance metrics that fail to accurately reflect the models' true ability to handle novel, unseen data, thereby undermining the validity of benchmark comparisons and the meaningful assessment of model capabilities.

In this paper, we present dynamic benchmarking\footnote{Released at: \url{https://www.kaggle.com/datasets/aspartic/dynamic-benchmark}}, a new benchmarking framework to address the aforementioned challenge.
Our framework takes an established code benchmark and mutates each test case with a set of semantic-preserving mutations. 
All the mutated code snippets are syntactically different from the original benchmark while having identical code semantics.
For instance, \F~\ref{fig:fig1_2} is a mutant of \F~\ref{fig:fig1_1} --- despite being syntactically different, both test cases have identical execution behaviors.
We instantiate our framework using a set of mutations that work on both the \textit{code syntax} and \textit{code structure} levels.
At the code syntax level, we focus on variable naming, \ie, normalizing or randomizing variable names, to remove irrelevant assistance information from them.\footnote{We \textit{do not} use adversarially mutated variable names to deliberately mislead models.}
At the code structure level, we focus on three core code structures, \ie, assignment, conditional branching, and loops.
By requiring models to process these mutated versions, we ensure that the evaluation accurately measures the model's capacity for code understanding and reasoning, rather than its ability to recall specific code patterns from its training data.

We evaluate our mutations on four benchmarks covering two coding tasks. The results show that in our new benchmarks, all models' performance declines significantly, up to 40\% lower than the original benchmarks. In order to evaluate whether or not our dynamic benchmarking approach can resist the data contamination problem, we fine-tuned a model with the original benchmark. The results show that our dynamically constructed benchmarks are moderately affected and are still useful in evaluating the fine-tuned model.


\vspace{-5pt}
\section{Methodology}
\vspace{-5pt}

\label{sec:methodology}

\subsection{Dynamic Benchmarking Framework}

LLM-based code reasoning generates outcomes based on high-level instructions and source code. Typically, the instructions describe the specific reasoning task in natural language, while the source code represents the program to be analyzed. Generally, we expect the LLM to derive results through reasoning based on the provided instructions and source code. 
Formally, for a given benchmark with $n$ test cases $\{<ins_i,p_i,o_i>\}, i\in[1,n]$ and a model $LLM$ under evaluation. 
The benchmark evaluates how correct $LLM(ins_i,p_i)$ is compared to the reference result $o_i$.
Ideally, all $\{<ins_i,p_i,o_i>\}$ should be unknown to the model. However, due to the data contamination problem, $\{<ins_i,p_i,o_i>\}$ are often leaked.

\begin{figure}[ht]
\centering
\includegraphics[width=0.8\columnwidth]{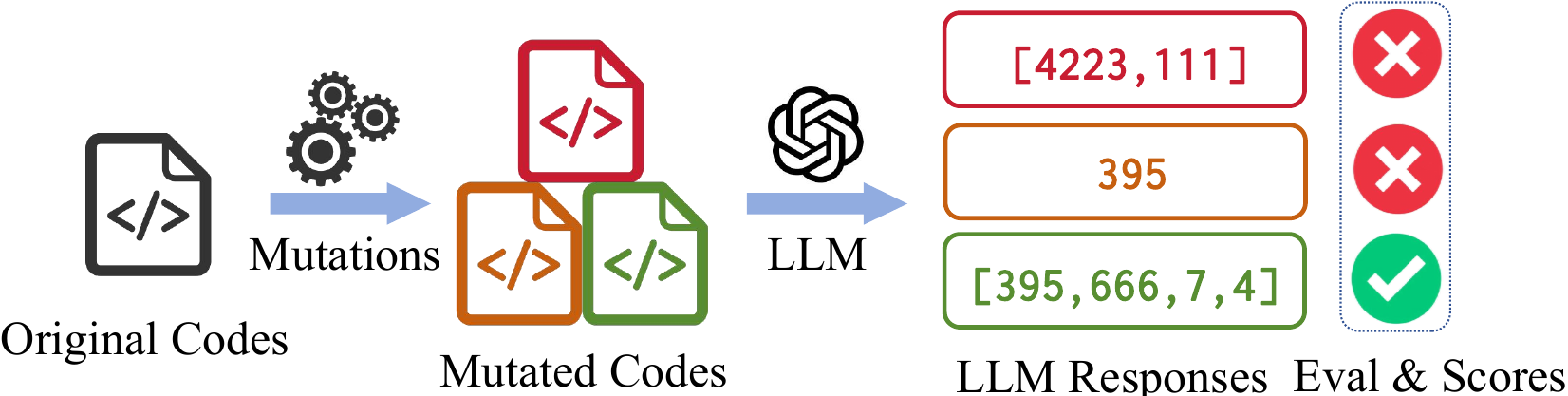}
\caption{The dynamic benchmarking framework.}
\label{fig:workflow}
\end{figure}

\F~\ref{fig:workflow} shows the high-level workflow of our dynamic benchmarking framework. Given one $<ins_i,p_i,o_i>$, our framework applies a set of semantics-preserving mutations $M_k$ to transform the test case into multiple ones as follows:
$$
<ins_i,p_i'=M_{k_1}(\cdots M_{k_{m}}(p_i)), o_i>,
$$
where $M_{k_{j}}$ is one of the available mutations. The new program $p_i'$ is sequentially mutated from the original program $p$. Since all $M_{k_{j}}$ are semantics-preserving, the final reference result is still $o_i$.
All the new test cases are dynamically generated from the existing benchmark and are then used for evaluating models.
To ensure the usability of newly generated test cases, each mutation method is governed by the following two fundamental requirements:


\textbf{\textit{Semantic Equivalence.}} The mutated program $p^\prime$ must maintain execution behavior identical to the original program $p$. Formally, for all possible inputs $x$ of program $p$, the following condition should be satisfied: 
$$\forall x, [[p]]_x = [[M_{k_j}(p)]]_x,$$
where $[[p]]_x$ denotes the execution result of program $p$ on the input $x$. This requirement guarantees \emph{the mutation preserves the program's semantics}.



\textbf{\textit{Syntactic Divergence.}} The mutated program $p^\prime$ should be syntactically different from the original program $p$ in a non-trivial manner. This property ensures that even if $p$ is included in the training set of a model, $p'$ can still be used to evaluate the model. 

In the following sections, we will propose a series of mutation methods at two levels: \textit{code syntax} and \textit{code structure}. 

\begin{figure*}[ht]
\begin{center}
\begin{tabular}{ccc}
\subfigure[Original Code]{
    \includegraphics[width=0.3\columnwidth]{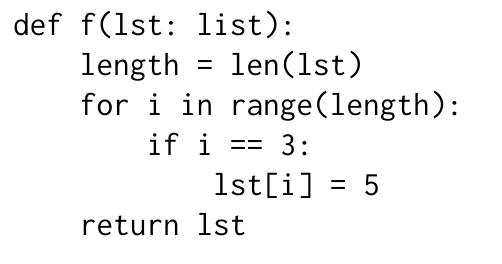}
    \label{fig:original} 
} &
\subfigure[Variable Normalization \uppercase\expandafter{\romannumeral1}]{
    \includegraphics[width=0.3\columnwidth]{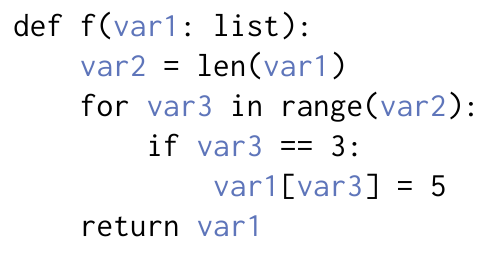}
    \label{fig:varnorm1} 
} &
\subfigure[Variable Normalization \uppercase\expandafter{\romannumeral2}]{
    \includegraphics[width=0.3\columnwidth]{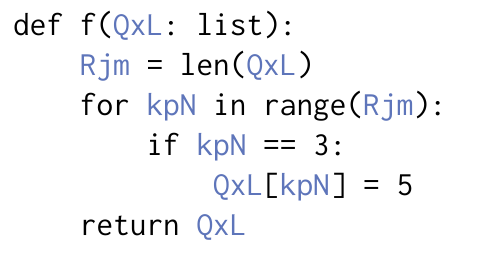}
    \label{fig:varnorm2} 
} \\
\subfigure[Constant Unfolding]{
    \raisebox{0.5cm}{\includegraphics[width=0.3\columnwidth]{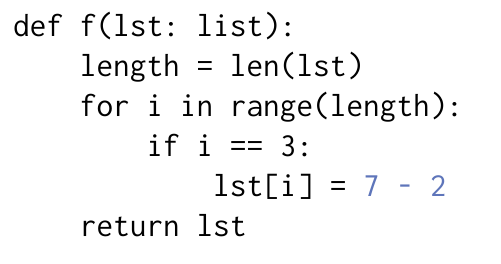}}
    \label{fig:constunfo} 
} &
\subfigure[For to While]{
    \includegraphics[width=0.3\columnwidth]{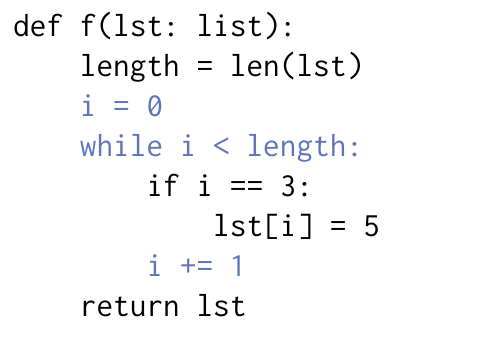}
    \label{fig:for2while} 
} &
\subfigure[Condition Augmentation]{
    \raisebox{0.3cm}{\includegraphics[width=0.3\columnwidth]{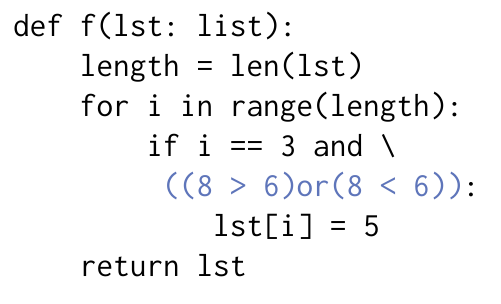}}
    \label{fig:condaug} 
} \\
\end{tabular}
\caption{Illustrative examples of our code syntax and code structure mutations. (a) is the code from the original benchmark, while (b)-(f) are new codes by applying one of our mutations.}
\label{fig:mutate} 
\vspace{-20pt}
\end{center}
\end{figure*}

\vspace{-5pt}
\subsection{Code Syntax Mutations}
In the process of models' reasoning about code, conventional static benchmarks typically employ meaningful variable names, which inadvertently provide models with additional cues that may influence their reasoning process. This phenomenon allows models to potentially utilize syntactic sugar information, particularly through variable naming patterns, as auxiliary features to facilitate code comprehension and reasoning. As demonstrated in \Cref{fig:fig1}, models can leverage semantically rich variable names to infer program functionality without fully grasping the underlying algorithmic logic. 
Moreover, once the static benchmark is leaked into the training set, variable names will become convenient objects that can be easily memorized.

\textbf{{Variable Normalization.}} To address these critical issues, we propose implementing \textit{Variable Normalization} as a methodological solution to build a dynamic benchmark. 
Specifically, we implement two distinct normalization schemes: \textbf{\textit{VarNormI}}, a sequential naming convention (e.g., \texttt{var1}, \texttt{var2}, \texttt{var3}, ...) that maintains variable uniqueness while eliminating semantic cues, and \textbf{\textit{VarNormII}}, a fixed-length randomized string generation system that ensures complete obfuscation of any potential naming-related information. 
\Cref{fig:original} presents a piece of original code with a format similar to CRUXEval. \Cref{fig:varnorm1} and \Cref{fig:varnorm2} are the new mutants after applying VarNormI and VarNormII mutations.
As the figure illustrates, variable normalization effectively eliminates the syntactic sugar information embedded in descriptive variable names such as \texttt{lists} and \texttt{length}.
By obfuscating the semantic information contained in variable names, this technique effectively forces LLMs to focus on the underlying structural and logical relationships within the code rather than relying on superficial naming patterns. Furthermore, since variable normalization exclusively modifies variable names without altering the underlying code structure, it inherently guarantees both semantic equivalence and syntactic divergence between the original and mutated code.

\vspace{-5pt}

\subsection{Code Structure Mutation}

\vspace{-5pt}
Understanding code involves more than just syntactic information. It also involves understanding how data moves through code (data flow) and how it makes decisions (control flow). 
To address data contamination in such code structures, we present a structural transformation approach for dynamic benchmark construction. Our approach systematically changes the structure of code samples while preserving their semantic equivalence and functional integrity. 
Specifically, we focus on three fundamental and ubiquitous code structures that represent core programming constructs: \textit{Assignment}, \textit{Loop}, and \textit{Branch}. We employ Python syntax as our demonstration framework to systematically illustrate the implementation of our mutation methodology. 

\textbf{{Assignment.}} Our analysis specifically targets the fundamental case of constant variable assignment, as exemplified by basic statements like \texttt{a = 5}. To address this, a mutation technique termed \textbf{\textit{Constant Unfolding}} has been developed, which systematically decomposes constants (e.g., integers) into equivalent arithmetic expressions while rigorously preserving semantic equivalence throughout the transformation process. 
This transformation is a reversion of the classical \textit{constant folding} optimization in compiler design~\cite{wirth1996compiler}.
As demonstrated in \Cref{fig:constunfo}, \textit{Constant Unfolding} transforms the assignment statement \texttt{lists[i] = 5} into \texttt{lists[i] = 7 - 2}. This mutation method only transforms an integer to an expression, so the semantic equivalence is maintained. Moreover, the proportion of text requiring modification remains minimal. By transforming straightforward integer assignments into equivalent arithmetic expressions, we create a subtle yet effective barrier against simple pattern recognition while preserving the essential reasoning challenges. 

\textbf{{Loop.}} In the domain of iterative control structures, two primary loop paradigms exist: \textit{for} loops and \textit{while} loops. These loop constructs, while syntactically distinct, achieve comparable computational complexity and can be mutually transformed through systematic restructuring. Our mutation framework incorporates the \textbf{\textit{For to While}} strategy, which transforms all \textit{for} loops into functionally equivalent \textit{while} loops and vice versa, as exemplified in \Cref{fig:for2while}. This mutation strategy also meets our requirements for mutation methods. Firstly, the semantic equivalence between \textit{for} loops and \textit{while} loops is guaranteed due to their interchangeable nature in representing iterative processes. This transformation maintains identical program behavior while introducing syntactic variations. Secondly, the modification scope is significantly limited, as only the loop header requires alteration.

\textbf{{Branch.}} Within conditional branching structures, we implement a transformation technique called \textbf{\textit{Condition Augmentation}}, which introduces tautological expressions into existing conditions while preserving the original logical outcome. 
At a high level, given an \textit{if} condition, it can always be equivalently transformed in the following way:
$$
if(C) \Leftrightarrow {if}(C\ \&\ \mbox{True} )  \Leftrightarrow {if}(C\ ||\ \mbox{False} )
$$
Guided by this principle, our implementation instantiates the condition \textit{True} or \textit{False} with predetermined yet complex expressions.
For example, as illustrated in \Cref{fig:condaug}, we augment the condition \texttt{i == 3} by incorporating a tautological expression \texttt{(8 > 6) or (8 < 6)}, resulting in the modified condition \texttt{i == 3 and ((8 > 6) or (8 < 6))} while maintaining the original logical equivalence. 
In addition, there are some optional tautology expressions, such as \texttt{True or False}. In actual experiments, we will use these tautology expressions in combination.
The mutation approach solely introduces tautological conditions in the branching part, leaving the code semantics unaffected. Thus, it meets our requirements.

{\color{black}{
\textbf{Discussion: The ``naturalness'' of mutated code.}
We deliberately exclude considerations of code ``naturalness'' in our evaluation framework for several reasons.
First, existing naturalness metrics, such as language model likelihoods ~\cite{khanfir2022codebertntcodenaturalnesscodebert, yang2024dependency}, are inherently biased towards their training data distributions and fail to capture the nuanced aspects of code quality, such as idiomatic usage, coding conventions, and style consistency.
More fundamentally, there exists no precise or universally accepted metric for evaluating code naturalness, making such assessments inherently subjective and potentially misleading.
This limitation is particularly relevant given that many benchmarks, such as CRUXEval's test functions which are generated using Code Llama, are deliberately constructed rather than derived from real-world code.
In this context, the pursuit of naturalness becomes a secondary concern. 
For code language models, the true value of a benchmark lies in its ability to rigorously evaluate a model's fundamental capabilities: its understanding of code semantics and its capacity for logical reasoning, rather than the superficial syntax.
}}

\vspace{-5pt}
\subsection{Multi-Mutation}
\vspace{-5pt}
All the aforementioned mutation methods operate independently of each other. As a result, these methods can be combined for use, enabling multi-mutation. This combined approach still satisfies our two key requirements for mutation methods. Firstly, since none of the individual mutations affect the program's execution outcome, multi-mutation likewise preserves the program's original behavior. Secondly, while individual mutations introduce minimal changes to the code text, the cumulative impact of multi-mutation remains limited and manageable. Therefore, this approach can be utilized effectively to build dynamic benchmarks and further test the code reasoning capabilities of LLMs.

\vspace{-5pt}
\section{Experimental Setup}
\vspace{-5pt}

Our evaluation aims to answer the following Research Questions (RQs):
\begin{enumerate}[leftmargin=15pt, topsep=0pt, noitemsep]
    \item (\textbf{Impact of single mutation}) How does models' performance change under each single mutation?
    \item (\textbf{Impact of multi-mutation}) How does models' performance change under multi-mutation? 
    \item (\textbf{Mitigation of data contamination}) Can dynamic benchmarking mitigate data contamination?
    \item (\textbf{Complexity of dynamic benchmarks}) How much additional complexity does our approach introduce to the original benchmarks?
\end{enumerate}


We selected a variety of both closed and open-source LLMs. Specifically, we chose models including GPT-4o mini~\cite{hurst2024gpt}, DeepSeek V3~\cite{deepseekai2024deepseekv3technicalreport}, Llama 3.1 series~\cite{grattafiori2024llama3herdmodels}, Qwen2.5-Coder series~\cite{hui2024qwen2}, and StarCoder2 series~\cite{lozhkov2024starcoder2stackv2}. Our models are sourced from the APIs provided by OpenRouter\footnote{\url{https://openrouter.ai/}} and the open-source models available on Hugging Face\footnote{\url{https://huggingface.co/}}.
In our experiments, we select two popular tasks: \textit{code execution} and \textit{code translation}. For code execution, we selected the CRUXEval benchmark~\cite{gu2024cruxeval}, while for code translation, we selected the Code Lingua benchmark (which includes two sub-datasets, Avatar and CodeNet)~\cite{pan2024lost} and the TransCoder benchmark~\cite{lachaux2020unsupervisedtranslationprogramminglanguages, yang2024exploringunleashingpowerlarge}, on which we conduct experiment on Python to Java translation.

For all LLMs, we set the temperature to 0.2 to evaluate their Pass@1 metric~\cite{chen2021evaluatinglargelanguagemodels}. We generate 5 results for each sample in the benchmarks. On the aforementioned four datasets, we applied five distinct mutation methods to each program within the datasets. These methods, as outlined in \Cref{sec:methodology}, include two types of Variable Normalization (\textbf{VarNormI} and \textbf{VarNormII}), Constant Unfolding (\textbf{ConstUnfold}), For to While (\textbf{For2While}), and Condition Augmentation (\textbf{CondAug}). 


As for multi-mutation, since we have five different mutation methods, there are dozens of combinations in theory. For a practical evaluation overhead, we randomly select three mutation combinations that cover all mutation methods from different categories: 
\textbf{FUV} (For2While (F) + ConstUnfold (U) + VarNormII (V)), \textbf{AUV} (CondAug (A) + ConstUnfold (U) + VarNormI (V)) and \textbf{AFU} (CondAug (A) + For2While (F) + ConstUnfold (U)).

\vspace{-5pt}
\section{Results and Analysis}
\vspace{-5pt}

\label{sec:result}


\subsection{RQ1: Impact of Single Mutation}

Due to space constraints, we present only the model Pass@1 performance results on both original static and our dynamic CRUXEval and CodeNet benchmarks in the main text. \Cref{tab:cnet_performance} shows the results. 

\begin{wrapfigure}{r}[0cm]{0pt}      
    \centering
    \includegraphics[width=0.5\textwidth]{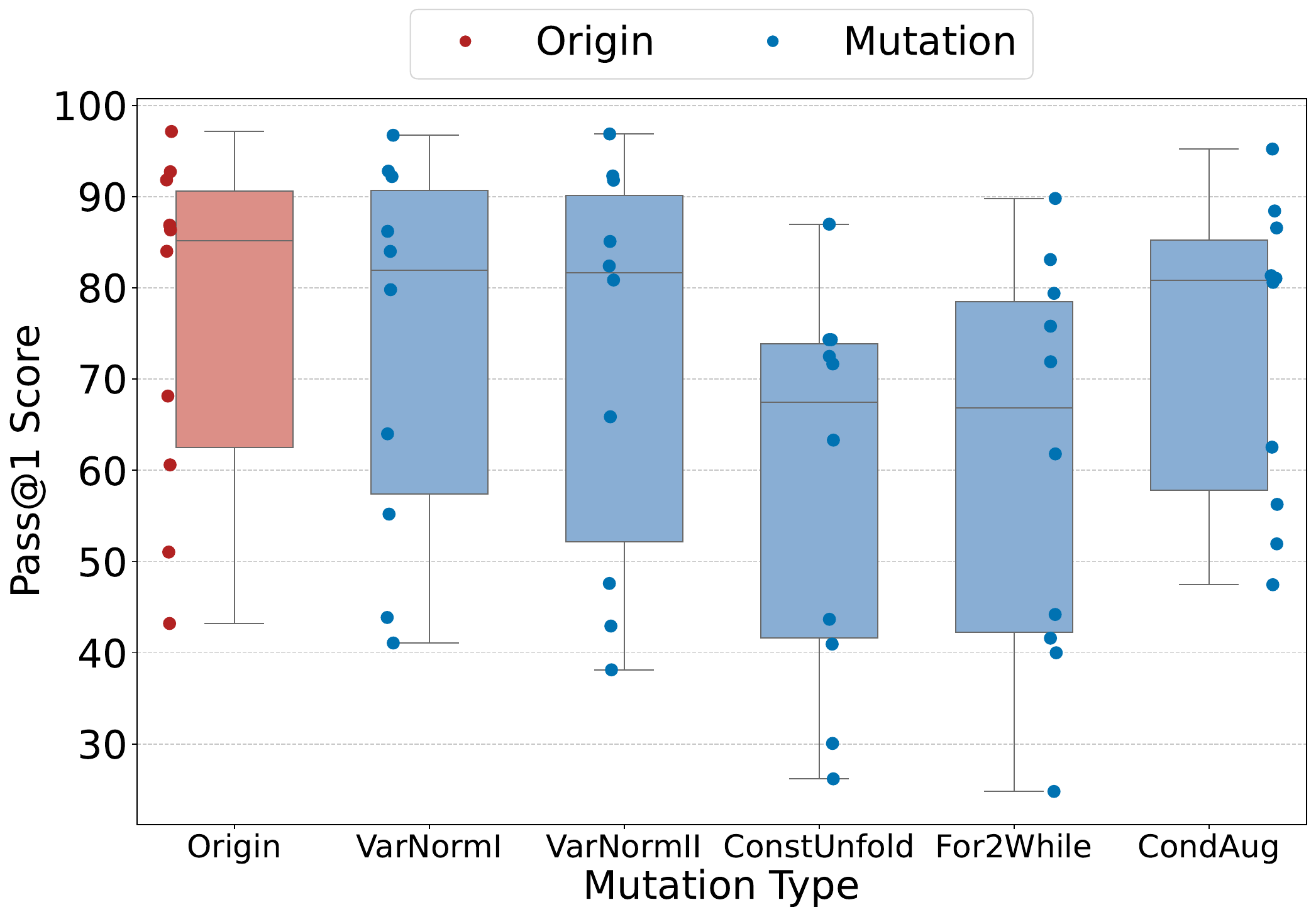} 
    \caption{Comparison of model performance distributions under different mutations of CodeNet.} 
    \label{fig:mutations_boxplot} 
\end{wrapfigure}
\textbf{\textit{General Tendency.}} The table reveals a consistent performance degradation tendency across most mutations, indicating that models demonstrate significantly reduced comprehension capabilities when processing mutated code compared to their performance on original code. For instance, after applying the Constant Unfolding method, the performance of many models dropped by more than 10\% compared to their performance on the original benchmark. This significant decline clearly demonstrates that models exhibit reduced code reasoning capabilities on dynamic benchmarks, thereby validating the effectiveness of our approach. 
Moreover, we observe that in a small number of cases, models exhibit slightly improved reasoning capabilities on the mutated benchmark compared to the original benchmark, particularly in the StarCoder2 series. This phenomenon may be attributed to the fact that the StarCoder2 models are not instruction-tuned, resulting in weaker adherence to instructions. As a consequence, their performance on the benchmark may not fully reflect their potential and is more susceptible to randomness, leading to such reversals.


\begin{table*}[htbp]
\centering
\setlength{\tabcolsep}{1pt}
\caption{Model Pass@1 performance on original and single mutated CRUXEval and CodeNet benchmarks. Performance drops of more than 10\% are highlighted.}
\label{tab:cnet_performance}
\begin{tabular}{l|l|c|lllll}
\toprule
\textbf{Type} & \textbf{Model} & \textbf{Origin} & \textbf{VarNormI} & \textbf{VarNormII} & \textbf{ConstUnfold} & \textbf{For2While} & \textbf{CondAug} \\

\midrule
\multirow{10}{*}{\rotatebox[origin=c]{90}{CRUXEval}}
& DeepSeek V3 & 65.80 & 63.91 {\color[rgb]{0.0,0.5,0.0}\scriptsize(-1.89)} & 63.01 {\color[rgb]{0.0,0.5,0.0}\scriptsize(-2.79)} & 46.70 {\color[rgb]{0.0,0.5,0.0}\scriptsize(\myhigh{-19.10})} & 58.01 {\color[rgb]{0.0,0.5,0.0}\scriptsize(-7.79)} & 63.07 {\color[rgb]{0.0,0.5,0.0}\scriptsize(-2.73)} \\
& GPT-4o mini & 53.60 & 51.80 {\color[rgb]{0.0,0.5,0.0}\scriptsize(-1.80)} & 52.08 {\color[rgb]{0.0,0.5,0.0}\scriptsize(-1.52)} & 34.51 {\color[rgb]{0.0,0.5,0.0}\scriptsize(\myhigh{-19.09})} & 47.61 {\color[rgb]{0.0,0.5,0.0}\scriptsize(-5.99)} & 54.71 {\color[rgb]{0.0,0.5,0.0}\scriptsize(-3.89)} \\
& Llama 3.1 70B & 53.20 & 51.21 {\color[rgb]{0.0,0.5,0.0}\scriptsize(-1.99)} & 51.94 {\color[rgb]{0.0,0.5,0.0}\scriptsize(-1.26)} & 34.95 {\color[rgb]{0.0,0.5,0.0}\scriptsize(\myhigh{-18.25})} & 45.95 {\color[rgb]{0.0,0.5,0.0}\scriptsize(-7.25)} & 53.10 {\color[rgb]{0.0,0.5,0.0}\scriptsize(-0.10)} \\
& Llama 3.1 8B & 31.20 & 30.10 {\color[rgb]{0.0,0.5,0.0}\scriptsize(-1.10)} & 28.46 {\color[rgb]{0.0,0.5,0.0}\scriptsize(-2.74)} & 21.80 {\color[rgb]{0.0,0.5,0.0}\scriptsize(-9.40)} & 20.46 {\color[rgb]{0.0,0.5,0.0}\scriptsize(\myhigh{-10.74})} & 23.96 {\color[rgb]{0.0,0.5,0.0}\scriptsize(-7.24)} \\
& Qwen2.5 32B & 65.50 & 63.59 {\color[rgb]{0.0,0.5,0.0}\scriptsize(-1.91)} & 63.02 {\color[rgb]{0.0,0.5,0.0}\scriptsize(-2.48)} & 40.26 {\color[rgb]{0.0,0.5,0.0}\scriptsize(\myhigh{-25.24})} & 57.22 {\color[rgb]{0.0,0.5,0.0}\scriptsize(-8.28)} & 60.43 {\color[rgb]{0.0,0.5,0.0}\scriptsize(-5.07)} \\
& Qwen2.5 14B & 58.30 & 57.27 {\color[rgb]{0.0,0.5,0.0}\scriptsize(-1.03)} & 57.10 {\color[rgb]{0.0,0.5,0.0}\scriptsize(-1.20)} & 39.21 {\color[rgb]{0.0,0.5,0.0}\scriptsize(\myhigh{-19.09})} & 53.01 {\color[rgb]{0.0,0.5,0.0}\scriptsize(-5.29)} & 58.98 {\color[rgb]{1.0,0.0,0.0}\scriptsize(+0.68)} \\
& Qwen2.5 7B & 45.70 & 43.77 {\color[rgb]{0.0,0.5,0.0}\scriptsize(-1.93)} & 43.97 {\color[rgb]{0.0,0.5,0.0}\scriptsize(-1.73)} & 26.55 {\color[rgb]{0.0,0.5,0.0}\scriptsize(\myhigh{-19.15})} & 42.03 {\color[rgb]{0.0,0.5,0.0}\scriptsize(-3.67)} & 42.14 {\color[rgb]{0.0,0.5,0.0}\scriptsize(-3.56)} \\
& StarCoder2 15B & 33.10 & 12.43 {\color[rgb]{0.0,0.5,0.0}\scriptsize(\myhigh{-20.67})} & 17.07 {\color[rgb]{0.0,0.5,0.0}\scriptsize(\myhigh{-16.03})} & 21.45 {\color[rgb]{0.0,0.5,0.0}\scriptsize(\myhigh{-11.65})} & 31.37 {\color[rgb]{0.0,0.5,0.0}\scriptsize(-1.73)} & 29.20 {\color[rgb]{0.0,0.5,0.0}\scriptsize(-3.90)} \\
& StarCoder2 7B & 33.80 & 24.64 {\color[rgb]{0.0,0.5,0.0}\scriptsize(-9.16)} & 25.27 {\color[rgb]{0.0,0.5,0.0}\scriptsize(-8.53)} & 20.84 {\color[rgb]{0.0,0.5,0.0}\scriptsize(\myhigh{-12.96})} & 33.27 {\color[rgb]{0.0,0.5,0.0}\scriptsize(-0.53)} & 32.94 {\color[rgb]{0.0,0.5,0.0}\scriptsize(-0.86)} \\
& StarCoder2 3B & 30.00 & 25.32 {\color[rgb]{0.0,0.5,0.0}\scriptsize(-4.68)} & 24.94 {\color[rgb]{0.0,0.5,0.0}\scriptsize(-5.06)} & 20.79 {\color[rgb]{0.0,0.5,0.0}\scriptsize(-9.21)} & 27.97 {\color[rgb]{0.0,0.5,0.0}\scriptsize(-2.03)} & 31.76 {\color[rgb]{1,0,0}\scriptsize(+1.76)} \\

\midrule
\multirow{10}{*}{\rotatebox[origin=c]{90}{CodeNet}}
& Deepseek V3 & 94.70 & 96.73 {\color[rgb]{1,0,0}\scriptsize(+2.03)} & 96.87 {\color[rgb]{1,0,0}\scriptsize(+2.17)} & 86.98 {\color[rgb]{0.0,0.5,0.0}\scriptsize(-7.72)} & 89.80 {\color[rgb]{0.0,0.5,0.0}\scriptsize(-4.90)} & 95.22 {\color[rgb]{1,0,0}\scriptsize(+0.52)} \\
& GPT-4o mini & 90.20 & 92.20 {\color[rgb]{1,0,0}\scriptsize(+2.00)} & 91.80 {\color[rgb]{1,0,0}\scriptsize(+1.60)} & 72.49 {\color[rgb]{0.0,0.5,0.0}\scriptsize(\myhigh{-17.71})} & 83.10 {\color[rgb]{0.0,0.5,0.0}\scriptsize(-7.10)} & 88.43 {\color[rgb]{0.0,0.5,0.0}\scriptsize(-1.77)} \\
& Llama 3.1 70B & 88.00 & 79.80 {\color[rgb]{0.0,0.5,0.0}\scriptsize(-8.20)} & 80.86 {\color[rgb]{0.0,0.5,0.0}\scriptsize(-7.14)} & 74.32 {\color[rgb]{0.0,0.5,0.0}\scriptsize(\myhigh{-13.68})} & 75.80 {\color[rgb]{0.0,0.5,0.0}\scriptsize(\myhigh{-12.20})} & 81.34 {\color[rgb]{0.0,0.5,0.0}\scriptsize(-6.66)} \\
& Llama 3.1 8B & 66.70 & 64.00 {\color[rgb]{0.0,0.5,0.0}\scriptsize(-2.70)} & 65.87 {\color[rgb]{0.0,0.5,0.0}\scriptsize(-0.83)} & 43.67 {\color[rgb]{0.0,0.5,0.0}\scriptsize(\myhigh{-23.03})} & 44.20 {\color[rgb]{0.0,0.5,0.0}\scriptsize(\myhigh{-22.50})} & 56.27 {\color[rgb]{0.0,0.5,0.0}\scriptsize(\myhigh{-10.43})} \\
& Qwen2.5 32B & 92.70 & 92.80 {\color[rgb]{1,0,0}\scriptsize(+0.10)} & 92.27 {\color[rgb]{0.0,0.5,0.0}\scriptsize(-0.43)} & 74.32 {\color[rgb]{0.0,0.5,0.0}\scriptsize(\myhigh{-18.38})} & 79.40 {\color[rgb]{0.0,0.5,0.0}\scriptsize(\myhigh{-13.30})} & 86.57 {\color[rgb]{0.0,0.5,0.0}\scriptsize(-6.13)} \\
& Qwen2.5 14B & 86.30 & 86.20 {\color[rgb]{0.0,0.5,0.0}\scriptsize(-0.10)} & 85.09 {\color[rgb]{0.0,0.5,0.0}\scriptsize(-1.21)} & 71.66 {\color[rgb]{0.0,0.5,0.0}\scriptsize(\myhigh{-14.64})} & 71.90 {\color[rgb]{0.0,0.5,0.0}\scriptsize(\myhigh{-14.40})} & 80.60 {\color[rgb]{0.0,0.5,0.0}\scriptsize(-5.70)} \\
& Qwen2.5 7B & 84.10 & 84.00 {\color[rgb]{0.0,0.5,0.0}\scriptsize(-0.10)} & 82.40 {\color[rgb]{0.0,0.5,0.0}\scriptsize(-1.70)} & 63.31 {\color[rgb]{0.0,0.5,0.0}\scriptsize(\myhigh{-20.79})} & 61.80 {\color[rgb]{0.0,0.5,0.0}\scriptsize(\myhigh{-22.30})} & 81.04 {\color[rgb]{0.0,0.5,0.0}\scriptsize(-3.06)} \\
& StarCoder2 15B & 58.60 & 55.20 {\color[rgb]{0.0,0.5,0.0}\scriptsize(-3.40)} & 47.60 {\color[rgb]{0.0,0.5,0.0}\scriptsize(\myhigh{-11.00})} & 40.95 {\color[rgb]{0.0,0.5,0.0}\scriptsize(\myhigh{-17.65})} & 41.60 {\color[rgb]{0.0,0.5,0.0}\scriptsize(\myhigh{-17.00})} & 62.54 {\color[rgb]{1,0,0}\scriptsize(+3.94)} \\
& StarCoder2 7B & 43.00 & 41.07 {\color[rgb]{0.0,0.5,0.0}\scriptsize(-1.93)} & 38.13 {\color[rgb]{0.0,0.5,0.0}\scriptsize(-4.87)} & 26.18 {\color[rgb]{0.0,0.5,0.0}\scriptsize(\myhigh{-16.82})} & 40.00 {\color[rgb]{0.0,0.5,0.0}\scriptsize(-3.00)} & 47.46 {\color[rgb]{1,0,0}\scriptsize(+4.46)} \\
& StarCoder2 3B & 50.70 & 43.87 {\color[rgb]{0.0,0.5,0.0}\scriptsize(-6.83)} & 42.93 {\color[rgb]{0.0,0.5,0.0}\scriptsize(-7.77)} & 30.06 {\color[rgb]{0.0,0.5,0.0}\scriptsize(\myhigh{-20.64})} & 24.80 {\color[rgb]{0.0,0.5,0.0}\scriptsize(\myhigh{-25.90})} & 51.94 {\color[rgb]{1,0,0}\scriptsize(+1.24)} \\
\bottomrule
\end{tabular}
\end{table*}

\textbf{\textit{Benchmark Effectiveness Enhancement.}} The significant performance drop of models on the dynamic benchmark also enhances the usability of certain benchmarks. For example, as shown in \Cref{tab:cnet_performance}, on the original CodeNet benchmark, the GPT-4o mini model achieves a performance of 90.20\%, indicating that the original CodeNet benchmark is relatively simple and no longer suitable for evaluating highly advanced models. However, after applying the Constant Unfolding mutation, the Pass@1 performance of GPT-4o mini drops to 72.49\%. This demonstrates that our method makes this benchmark useful again by reintroducing meaningful differentiation among state-of-the-art models.

\textbf{\textit{Performance Differentiation.}} Moreover, the dynamic benchmark enhances the differentiation of performance among various models by applying mutations to the benchmark. \Cref{fig:mutations_boxplot} presents box plots illustrating the performance distribution of different models on the CodeNet benchmark before and after mutation. As shown in the figure, the performance distribution of models after mutation is significantly more spread out compared to the original static benchmark, which facilitates a better distinction between the capabilities of different models. This is also evident in specific examples. On the original CodeNet benchmark, the DeepSeek V3 model achieves a performance of 94.70\%, while the GPT-4o mini model reaches 90.20\%, indicating a relatively small gap between the two. However, after applying the Constant Unfolding mutation, the Pass@1 performance of DeepSeek V3 drops to 86.98\%, whereas GPT-4o mini's performance declines significantly to 72.49\%, resulting in a much larger gap. This demonstrates that the dynamic benchmark enables a more effective evaluation of the performance differences among models.

\begin{table*}[htbp]
\centering
\setlength{\tabcolsep}{9pt}
\small
\caption{Model Pass@1 performance on original and multiple mutated CRUXEval and CodeNet benchmarks. Performance drops of more than 20\% are highlighted.}
\label{tab:multi_performance}
\begin{tabular}{l|l|c|lll}
\toprule
\textbf{Type} & \textbf{Model} & \textbf{Origin} & \textbf{FUV} & \textbf{AUV} & \textbf{AFU} \\
\midrule
\multirow{10}{*}{\rotatebox[]{0}{CRUXEval}}
& DeepSeek V3 & 65.80 & 45.93 {\color[rgb]{0.0,0.5,0.0}\scriptsize(-19.87)} & 52.21 {\color[rgb]{0.0,0.5,0.0}\scriptsize(-13.59)} & 50.63 {\color[rgb]{0.0,0.5,0.0}\scriptsize(-15.17)} \\
& GPT-4o mini & 53.60 & 31.28 {\color[rgb]{0.0,0.5,0.0}\scriptsize(\myhigh{-22.32})} & 42.18 {\color[rgb]{0.0,0.5,0.0}\scriptsize(-11.42)} & 37.94 {\color[rgb]{0.0,0.5,0.0}\scriptsize(-15.66)} \\
& Llama 3.1 70B & 53.20 & 36.79 {\color[rgb]{0.0,0.5,0.0}\scriptsize(-16.41)} & 40.89 {\color[rgb]{0.0,0.5,0.0}\scriptsize(-12.31)} & 37.57 {\color[rgb]{0.0,0.5,0.0}\scriptsize(-15.63)} \\
& Llama 3.1 8B & 31.20 & 15.93 {\color[rgb]{0.0,0.5,0.0}\scriptsize(-15.27)} & 23.26 {\color[rgb]{0.0,0.5,0.0}\scriptsize(-7.94)} & 20.48 {\color[rgb]{0.0,0.5,0.0}\scriptsize(-10.72)} \\
& Qwen2.5-Coder 32B & 65.50 & 37.25 {\color[rgb]{0.0,0.5,0.0}\scriptsize(\myhigh{-28.25})} & 47.82 {\color[rgb]{0.0,0.5,0.0}\scriptsize(-17.68)} & 39.58 {\color[rgb]{0.0,0.5,0.0}\scriptsize(\myhigh{-25.92})} \\
& Qwen2.5-Coder 14B & 58.30 & 33.90 {\color[rgb]{0.0,0.5,0.0}\scriptsize(\myhigh{-24.40})} & 47.71 {\color[rgb]{0.0,0.5,0.0}\scriptsize(-10.59)} & 39.15 {\color[rgb]{0.0,0.5,0.0}\scriptsize(-19.15)} \\
& Qwen2.5-Coder 7B & 45.70 & 31.61 {\color[rgb]{0.0,0.5,0.0}\scriptsize(-14.09)} & 32.02 {\color[rgb]{0.0,0.5,0.0}\scriptsize(-13.68)} & 31.96 {\color[rgb]{0.0,0.5,0.0}\scriptsize(-13.74)} \\
& StarCoder2 15B & 33.10 & 12.07 {\color[rgb]{0.0,0.5,0.0}\scriptsize(\myhigh{-21.03})} & 8.57 {\color[rgb]{0.0,0.5,0.0}\scriptsize(\myhigh{-24.53})} & 28.36 {\color[rgb]{0.0,0.5,0.0}\scriptsize(-4.74)} \\
& StarCoder2 7B & 33.80 & 20.33 {\color[rgb]{0.0,0.5,0.0}\scriptsize(-13.47)} & 20.05 {\color[rgb]{0.0,0.5,0.0}\scriptsize(-13.75)} & 32.49 {\color[rgb]{0.0,0.5,0.0}\scriptsize(-1.31)} \\
& StarCoder2 3B & 30.00 & 16.59 {\color[rgb]{0.0,0.5,0.0}\scriptsize(-13.41)} & 20.11 {\color[rgb]{0.0,0.5,0.0}\scriptsize(-9.89)} & 25.19 {\color[rgb]{0.0,0.5,0.0}\scriptsize(-4.81)} \\
\midrule
\multirow{10}{*}{\rotatebox[]{0}{CodeNet}}
& DeepSeek V3 & 94.70 & 78.93 {\color[rgb]{0.0,0.5,0.0}\scriptsize(-15.77)} & 89.68 {\color[rgb]{0.0,0.5,0.0}\scriptsize(-5.02)} & 78.54 {\color[rgb]{0.0,0.5,0.0}\scriptsize(-16.16)} \\
& GPT-4o mini & 90.20 & 70.71 {\color[rgb]{0.0,0.5,0.0}\scriptsize(-19.49)} & 75.48 {\color[rgb]{0.0,0.5,0.0}\scriptsize(-14.72)} & 56.58 {\color[rgb]{0.0,0.5,0.0}\scriptsize(\myhigh{-33.62})} \\
& Llama 3.1 70B & 88.00 & 43.21 {\color[rgb]{0.0,0.5,0.0}\scriptsize(\myhigh{-44.79})} & 74.41 {\color[rgb]{0.0,0.5,0.0}\scriptsize(-13.59)} & 53.17 {\color[rgb]{0.0,0.5,0.0}\scriptsize(\myhigh{-34.83})} \\
& Llama 3.1 8B & 66.70 & 25.00 {\color[rgb]{0.0,0.5,0.0}\scriptsize(\myhigh{-41.70})} & 30.11 {\color[rgb]{0.0,0.5,0.0}\scriptsize(\myhigh{-36.59})} & 16.10 {\color[rgb]{0.0,0.5,0.0}\scriptsize(\myhigh{-50.60})} \\
& Qwen2.5-Coder 32B & 92.70 & 58.21 {\color[rgb]{0.0,0.5,0.0}\scriptsize(\myhigh{-34.49})} & 63.66 {\color[rgb]{0.0,0.5,0.0}\scriptsize(\myhigh{-29.04})} & 51.95 {\color[rgb]{0.0,0.5,0.0}\scriptsize(\myhigh{-40.75})} \\
& Qwen2.5-Coder 14B & 86.30 & 58.93 {\color[rgb]{0.0,0.5,0.0}\scriptsize(\myhigh{-27.37})} & 69.89 {\color[rgb]{0.0,0.5,0.0}\scriptsize(-16.41)} & 50.00 {\color[rgb]{0.0,0.5,0.0}\scriptsize(\myhigh{-36.30})} \\
& Qwen2.5-Coder 7B & 84.10 & 40.71 {\color[rgb]{0.0,0.5,0.0}\scriptsize(\myhigh{-43.39})} & 73.55 {\color[rgb]{0.0,0.5,0.0}\scriptsize(-10.55)} & 40.24 {\color[rgb]{0.0,0.5,0.0}\scriptsize(\myhigh{-43.86})} \\
& StarCoder2 15B & 58.60 & 32.86 {\color[rgb]{0.0,0.5,0.0}\scriptsize(\myhigh{-25.74})} & 56.34 {\color[rgb]{0.0,0.5,0.0}\scriptsize(-2.26)} & 25.12 {\color[rgb]{0.0,0.5,0.0}\scriptsize(\myhigh{-33.48})} \\
& StarCoder2 7B & 43.00 & 18.21 {\color[rgb]{0.0,0.5,0.0}\scriptsize(\myhigh{-24.79})} & 35.05 {\color[rgb]{0.0,0.5,0.0}\scriptsize(-7.95)} & 16.34 {\color[rgb]{0.0,0.5,0.0}\scriptsize(\myhigh{-26.66})} \\
& StarCoder2 3B & 50.70 & 19.29 {\color[rgb]{0.0,0.5,0.0}\scriptsize(\myhigh{-31.41})} & 43.01 {\color[rgb]{0.0,0.5,0.0}\scriptsize(-7.69)} & 19.27 {\color[rgb]{0.0,0.5,0.0}\scriptsize(\myhigh{-31.43})} \\
\bottomrule
\end{tabular}
\end{table*}

\vspace{-5pt}
\subsection{RQ2: Impact of Multi-Mutation}
\vspace{-5pt}

Here, we present a comparison between static and dynamic CRUXEval and CodeNet benchmarks based on multi-mutation, as shown in \Cref{tab:multi_performance}. By combining different mutation methods, the results show some new variations compared to using a single mutation method. We will explain these changes in detail below.

\vspace{-5pt}
\textbf{\textit{Drastic Performance Drop.}} From the table, it can be observed that the performance drop caused by multi-mutation is more significant than that of a single mutation. On CRUXEval, the performance drop can reach \textit{as high as 20\% or even close to 30\%}. On CodeNet, many models incur \textit{30\% to 50\%} performance drop.
These phenomena indicate that our method achieves even better results when applied in combination.
\vspace{-5pt}

\textbf{\textit{Ranking Changes.}} In \Cref{tab:cnet_performance}, although the absolute performance of the models has declined, the relative ranking changes are not significant. However, after applying multi-mutation, the relative rankings of the models have undergone observable changes. For example, in the CodeNet benchmark, the rankings of the three models from the Qwen2.5-Coder series experienced significant changes. The original ranking \textit{is 32B \textgreater 14B \textgreater 7B}, but after applying the FUV method, the ranking \textit{becomes 14B \textgreater 32B \textgreater 7B}, and with the AUV method, the ranking \textit{shifts to 7B \textgreater 14B \textgreater 32B}. These ranking changes also suggest that, when using the original static benchmark, the inference performance of these models might not have been accurately evaluated due to issues like data contamination. Additionally, the CodeNet Benchmark was released in January 2024, while the data collection for Qwen2.5-Coder was completed by February 2024. Our results indicate that some models may be implicitly affected by the data contamination issue.

\subsection{RQ3: Mitigation of Data Contamination}

\begin{wrapfigure}{r}[0cm]{0pt}    
    \centering
    \includegraphics[width=0.45\textwidth]{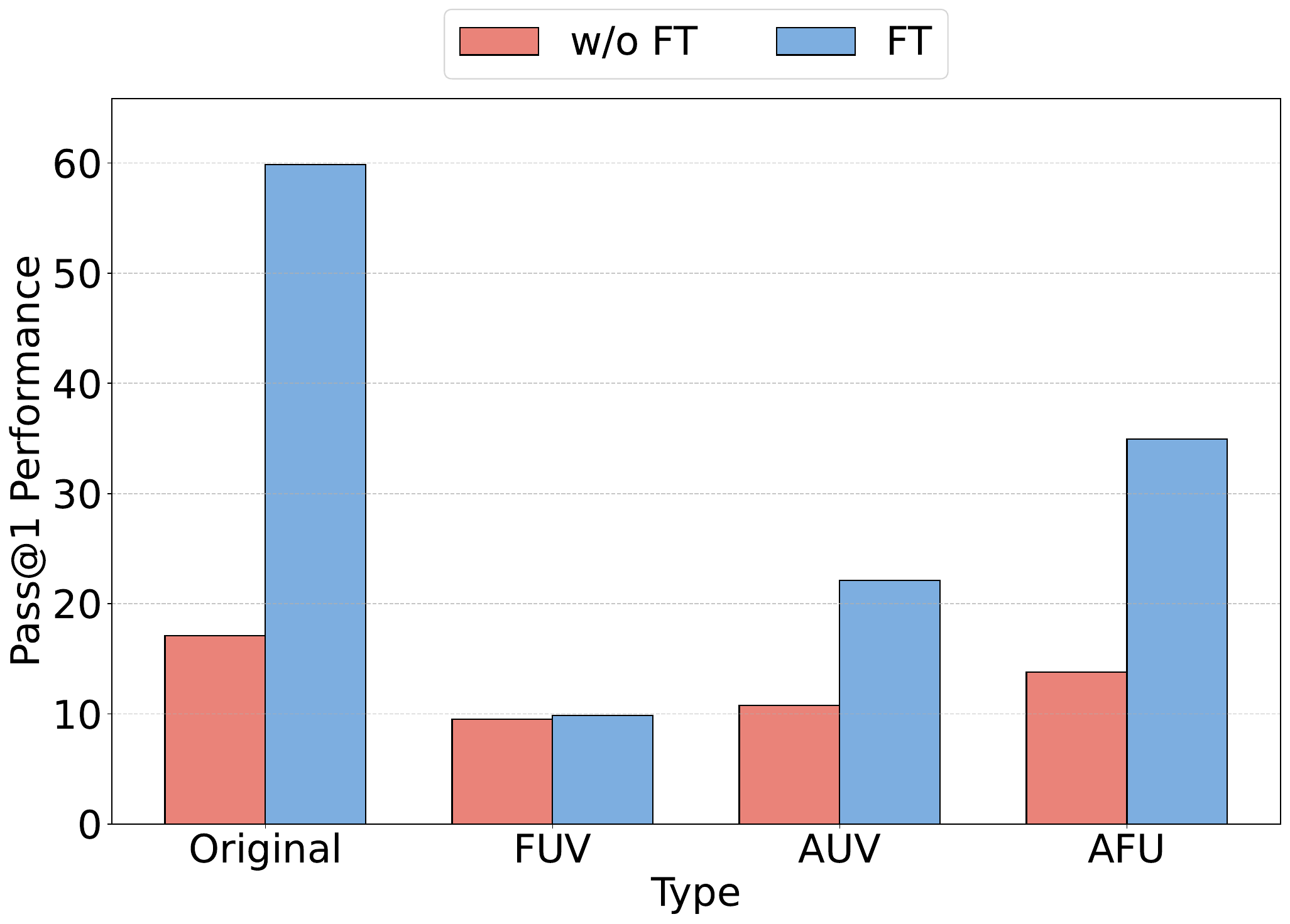} 
    \caption{Comparison of Pass@1 scores on static and dynamic CRUXEval, with and without(w/o) fine-tuning the model.}
    \label{fig:ft_performance} 
\end{wrapfigure}

In this section, we further illustrate the significant impact of data contamination and how our method can effectively mitigate these issues. 
{\color{black}{In RQ1 and RQ2, we find that the performance trends of different models under our dynamic benchmark are generally consistent. Due to computational resource constraints, in this section, we select the \texttt{Qwen2.5-Coder 1.5B Instruct} model and fine-tune it using the CRUXEval dataset to simulate the data contamination issue. In this experiment, we use a learning rate of 1e-4, a batch size of 2, and 20 training epochs. }}
We then evaluate the model on both the original static CRUXEval benchmark and our dynamic benchmark.
Since the previous evaluation shows the strong capability of multi-mutation, we use multi-mutation as the dynamic benchmarking approach. The results are shown in \Cref{fig:ft_performance}.

It shows that the fine-tuned model achieves approximately three times higher Pass@1 scores than the original model, indicating that it has memorized the dataset. This demonstrates that data contamination can severely undermine the ability of traditional static benchmarks to accurately reflect a model's true code reasoning capabilities.  
Moreover, when the fine-tuned model is evaluated on the dynamic benchmark, the Pass@1 scores also decline compared to those on the static benchmark. 
Particularly for the FUV method, \textit{the impact of fine-tuning on its evaluation is almost negligible}.
This effectively demonstrates that our approach maintains its efficacy in assessing the model's genuine reasoning capabilities, even in scenarios where test data has been incorporated into the training dataset, thereby meeting the criteria for syntactic divergence.

\subsection{RQ4: Complexity of Dynamic Benchmarks}

\begin{wraptable}{r}[-5pt]{0.5\textwidth}
\setlength{\tabcolsep}{3pt}
\vspace{-15pt}
\caption{BLEU scores of different mutated benchmarks compared to the original benchmarks.}
\label{tab:bleu}
\begin{tabular}{lccccc}
\toprule
\textbf{Benchmark} & \textbf{VI} & \textbf{VII} & \textbf{CU} & \textbf{F2W} & \textbf{CA} \\ 
\midrule
CRUXEval & 43.35 & 40.70 & 70.41 & 44.45 & 55.64 \\ 
Avatar & 50.28 & 50.22 & 70.36 & 63.19 & 72.43 \\ 
CodeNet & 56.46 & 56.91 & 69.55 & 59.92 & 67.45 \\ 
TransCoder & 37.53 & 37.50 & 61.03 & 52.21 & 64.93 \\ 
\bottomrule
\end{tabular}
\end{wraptable}
In \Cref{tab:bleu}, we compare the similarity between benchmarks obtained by different mutated methods and original benchmarks, using the BLEU score~\cite{papineni2002bleu} to measure the similarity. The results show that the BLEU score of the vast majority of dynamic benchmarks compared to the original benchmarks reaches \textit{above 50}, indicating that our method effectively reduced the model's reasoning performance without significantly altering benchmarks.

\vspace{-5pt}
\section{Related Work}
\vspace{-5pt}

\subsection{Code Reasoning}
\vspace{-5pt}

In the domain of code, various benchmarks are developed to evaluate the code reasoning abilities of LLMs.
For example, CRUXEval~\cite{gu2024cruxeval} concentrates on code execution challenges by supplying LLMs with a function and asking them either to produce outputs corresponding to given inputs or to identify a set of inputs that would yield a specified output. Meanwhile, benchmarks such as Code Lingua~\cite{pan2024lost} and TransCoder~\cite{lachaux2020unsupervisedtranslationprogramminglanguages, yang2024exploringunleashingpowerlarge} are geared towards code translation, aiming to assess LLMs' ability to convert code from one language to another. Several other benchmarks have also been developed to assess the code reasoning capabilities of LLMs across diverse tasks. For instance, QuixBugs~\cite{lin2017quixbugs} evaluates models' proficiency in program repair, while CoDesc~\cite{hasan-etal-2021-codesc} and similar benchmarks focus on assessing models' ability to code summarization.
\vspace{-5pt}

\subsection{Mutation Testing}
\vspace{-5pt}

Mutation testing has been extensively studied as a powerful technique for evaluating the effectiveness of test suites and improving software quality. 
A representative example can be found in compiler testing, where mutation testing of input programs has been effectively employed to identify subtle bugs and optimization issues in modern compilers~\cite{le2014compiler, sun2016finding,Shaohua10.1145/3656386}. 
In the domain of LLM testing, 
\citet{hooda2024large} conduct a comparative analysis of LLMs' performance on code generation tasks using both original and mutated datasets, revealing that current models demonstrate a limited understanding of fundamental programming concepts such as data flow and control flow. 
{\color{black}{Moreover, \citet{chen2025dynamicbenchmarkingreasoningcapabilities} enhances the evaluation accuracy of large language models by modifying the meaning and context of the natural language descriptions in programming problems.}}
\vspace{-5pt}

\subsection{Benchmark Reliability}
\vspace{-5pt}

Benchmark reliability has been a persistent issue throughout the development of deep learning, especially since LLMs became mainstream in the academic community. The remarkable generalization capabilities and task-handling abilities of LLMs have led to the creation of numerous benchmarks tailored to various tasks. However, the quality of these benchmarks varies significantly, and prior research has already highlighted this problem. For example, \citet{gulati2024putnam} proposed Putnam-AXIOM, demonstrating that current large models still perform poorly on mathematical problems. Similarly, \citet{liu2024your} introduced EvalPlus, which rigorously evaluates LLMs' code generation capabilities through mutation-based methods. Our work further advances LLM evaluation by focusing on code reasoning tasks, thereby contributing to a more comprehensive assessment system.



\vspace{-5pt}

\section{Conclusion}

\vspace{-5pt}

In this paper, we tackle a critical challenge in model evaluation: how to keep code benchmarks meaningful when models might have already seen them during training. Our solution, dynamic benchmarking, automatically transforms test programs while keeping their semantics intact.
Our experiments show this approach works remarkably well --- models struggle more with our transformed benchmarks, and even when a model is fine-tuned on the original benchmark, our dynamic versions still provide reliable evaluation.
This opens up new possibilities for keeping model evaluation fair and meaningful as models continue to advance.

\bibliographystyle{plainnat}
\bibliography{pymutate}







\appendix

\section{Benchmark Test Case Counts}
\Cref{tab:benchsize} enumerates the number of test cases in each benchmark and mutation method. Note that not all test cases can be mutated by each method. For example, out of all 800 test cases in CRUXEval, there are 306 cases containing \textit{for} loops and thus can be mutated by \textit{For2While}. 
In our evaluation, we only count the mutated test cases when evaluating each mutation method.

\begin{table}[htbp]
    \centering
    \caption{\#test cases in the original benchmarks and \#test cases that can be mutated by each mutation method.}
    \label{tab:benchsize}
    \setlength{\tabcolsep}{3pt} 
    \begin{tabular}{lccccc}
        \toprule
        Mutation  & CRUXEval & Avatar & CodeNet & TransCoder \\
        \midrule
        Original & 800 & 250 & 200 & 568 \\
        VarNormI & 785 & 131 & 150 & 568 \\
        VarNormII & 785 & 131 & 150 & 568 \\
        ConstUnfold & 455 & 239 & 169 & 546 \\
        For2While & 306 & 181 & 100 & 349 \\
        CondAug & 374 & 191 & 134 & 396 \\
        \bottomrule
    \end{tabular}
\end{table}

\section{Detailed Experimental Results}
In this section, we present all of our experimental results, which generally align with the trends reported in \Cref{sec:result}.


\subsection{More Results on Single Mutated Experiments}
More experimental results on single mutated benchmark are displayed in \Cref{tab:ava_performance}.

\begin{table*}[htbp]
\centering
\setlength{\tabcolsep}{1pt}
\caption{Model Pass@1 performance on original and single mutated Avatar and TransCoder benchmarks. Performance drops of more than 10\% are highlighted.}
\label{tab:ava_performance}
\begin{tabular}{l|l|c|lllll}
\toprule
\textbf{Type} & \textbf{Model} & \textbf{Origin} & \textbf{VarNormI} & \textbf{VarNormII} & \textbf{ConstUnfold} & \textbf{For2While} & \textbf{CondAug} \\

\midrule
\multirow{10}{*}{\rotatebox[origin=c]{90}{Avatar}}
& DeepSeek V3 & 80.64 & 85.42 {\color[rgb]{1,0,0}\scriptsize(+4.78)} & 81.98 {\color[rgb]{1,0,0}\scriptsize(+1.34)} & 72.68 {\color[rgb]{0.0,0.5,0.0}\scriptsize(-7.96)} & 73.48 {\color[rgb]{0.0,0.5,0.0}\scriptsize(-7.16)} & 81.68 {\color[rgb]{1,0,0}\scriptsize(+1.04)} \\
& GPT-4o mini & 72.48 & 82.13 {\color[rgb]{1,0,0}\scriptsize(+9.65)} & 74.05 {\color[rgb]{1,0,0}\scriptsize(+1.57)} & 54.77 {\color[rgb]{0.0,0.5,0.0}\scriptsize(\myhigh{-17.71})} & 67.90 {\color[rgb]{0.0,0.5,0.0}\scriptsize(-4.58)} & 73.25 {\color[rgb]{1,0,0}\scriptsize(+0.77)} \\
& Llama 3.1 70B & 72.80 & 64.35 {\color[rgb]{0.0,0.5,0.0}\scriptsize(\myhigh{-8.45})} & 72.6 {\color[rgb]{0.0,0.5,0.0}\scriptsize(-0.20)} & 57.87 {\color[rgb]{0.0,0.5,0.0}\scriptsize(\myhigh{-14.93})} & 62.71 {\color[rgb]{0.0,0.5,0.0}\scriptsize(\myhigh{-10.09})} & 70.16 {\color[rgb]{0.0,0.5,0.0}\scriptsize(-2.64)} \\
& Llama 3.1 8B & 46.48 & 49.01 {\color[rgb]{1,0,0}\scriptsize(+2.53)} & 52.98 {\color[rgb]{1,0,0}\scriptsize(+6.50)} & 29.87 {\color[rgb]{0.0,0.5,0.0}\scriptsize(\myhigh{-16.61})} & 36.24 {\color[rgb]{0.0,0.5,0.0}\scriptsize(\myhigh{-10.24})} & 43.04 {\color[rgb]{0.0,0.5,0.0}\scriptsize(-3.44)} \\
& Qwen2.5 32B & 75.64 & 78.76 {\color[rgb]{1,0,0}\scriptsize(+3.12)} & 76.64 {\color[rgb]{1,0,0}\scriptsize(+1.00)} & 59.42 {\color[rgb]{0.0,0.5,0.0}\scriptsize(\myhigh{-16.22})} & 65.08 {\color[rgb]{0.0,0.5,0.0}\scriptsize(\myhigh{-10.56})} & 70.68 {\color[rgb]{0.0,0.5,0.0}\scriptsize(-4.96)} \\
& Qwen2.5 14B & 65.20 & 70.08 {\color[rgb]{1,0,0}\scriptsize(+4.88)} & 70.53 {\color[rgb]{1,0,0}\scriptsize(+5.33)} & 54.31 {\color[rgb]{0.0,0.5,0.0}\scriptsize(\myhigh{-10.89})} & 57.46 {\color[rgb]{0.0,0.5,0.0}\scriptsize(-7.74)} & 64.08 {\color[rgb]{0.0,0.5,0.0}\scriptsize(-1.12)} \\
& Qwen2.5 7B & 64.60 & 61.83 {\color[rgb]{0.0,0.5,0.0}\scriptsize(-2.77)} & 65.80 {\color[rgb]{1,0,0}\scriptsize(+1.20)} & 43.18 {\color[rgb]{0.0,0.5,0.0}\scriptsize(\myhigh{-21.42})} & 51.27 {\color[rgb]{0.0,0.5,0.0}\scriptsize(\myhigh{-13.33})} & 58.12 {\color[rgb]{0.0,0.5,0.0}\scriptsize(-6.48)} \\
& StarCoder2 15B & 58.12 & 43.66 {\color[rgb]{0.0,0.5,0.0}\scriptsize(\myhigh{-14.46})} & 45.95 {\color[rgb]{0.0,0.5,0.0}\scriptsize(\myhigh{-12.17})} & 25.86 {\color[rgb]{0.0,0.5,0.0}\scriptsize(\myhigh{-32.26})} & 26.52 {\color[rgb]{0.0,0.5,0.0}\scriptsize(\myhigh{-31.60})} & 30.68 {\color[rgb]{0.0,0.5,0.0}\scriptsize(\myhigh{-27.44})} \\
& StarCoder2 7B & 40.00 & 30.84 {\color[rgb]{0.0,0.5,0.0}\scriptsize(-9.16)} & 28.85 {\color[rgb]{0.0,0.5,0.0}\scriptsize(\myhigh{-11.15})} & 16.40 {\color[rgb]{0.0,0.5,0.0}\scriptsize(\myhigh{-23.60})} & 20.88 {\color[rgb]{0.0,0.5,0.0}\scriptsize(\myhigh{-19.12})} & 27.54 {\color[rgb]{0.0,0.5,0.0}\scriptsize(\myhigh{-12.46})} \\
& StarCoder2 3B & 41.80 & 32.06 {\color[rgb]{0.0,0.5,0.0}\scriptsize(-9.74)} & 32.37 {\color[rgb]{0.0,0.5,0.0}\scriptsize(-9.43)} & 17.49 {\color[rgb]{0.0,0.5,0.0}\scriptsize(\myhigh{-24.31})} & 17.90 {\color[rgb]{0.0,0.5,0.0}\scriptsize(\myhigh{-23.90})} & 23.14 {\color[rgb]{0.0,0.5,0.0}\scriptsize(\myhigh{-18.66})} \\
\midrule
\multirow{10}{*}{\rotatebox[origin=c]{90}{TransCoder}}
& Deepseek V3 & 79.36 & 75.84 {\color[rgb]{0.0,0.5,0.0}\scriptsize(-3.52)} & 74.69 {\color[rgb]{0.0,0.5,0.0}\scriptsize(-4.67)} & 69.73 {\color[rgb]{0.0,0.5,0.0}\scriptsize(-9.63)} & 65.39 {\color[rgb]{0.0,0.5,0.0}\scriptsize(\myhigh{-13.97})} & 67.95 {\color[rgb]{0.0,0.5,0.0}\scriptsize(\myhigh{-11.41})} \\
& GPT-4o mini & 78.11 & 72.28 {\color[rgb]{0.0,0.5,0.0}\scriptsize(-5.83)} & 71.82 {\color[rgb]{0.0,0.5,0.0}\scriptsize(-6.29)} & 69.56 {\color[rgb]{0.0,0.5,0.0}\scriptsize(-8.55)} & 82.96 {\color[rgb]{1,0,0}\scriptsize(+4.85)} & 70.95 {\color[rgb]{0.0,0.5,0.0}\scriptsize(-7.16)} \\
& Llama 3.1 70B & 78.88 & 73.36 {\color[rgb]{0.0,0.5,0.0}\scriptsize(-5.52)} & 72.57 {\color[rgb]{0.0,0.5,0.0}\scriptsize(-6.31)} & 52.69 {\color[rgb]{0.0,0.5,0.0}\scriptsize(\myhigh{-26.19})} & 61.55 {\color[rgb]{0.0,0.5,0.0}\scriptsize(\myhigh{-17.33})} & 64.82 {\color[rgb]{0.0,0.5,0.0}\scriptsize(\myhigh{-14.06})} \\
& Llama 3.1 8B & 72.56 & 62.66 {\color[rgb]{0.0,0.5,0.0}\scriptsize(-9.90)} & 64.23 {\color[rgb]{0.0,0.5,0.0}\scriptsize(-8.33)} & 59.04 {\color[rgb]{0.0,0.5,0.0}\scriptsize(\myhigh{-13.52})} & 77.80 {\color[rgb]{1,0,0}\scriptsize(+5.24)} & 64.55 {\color[rgb]{0.0,0.5,0.0}\scriptsize(-8.01)} \\
& Qwen2.5 32B & 73.24 & 55.48 {\color[rgb]{0.0,0.5,0.0}\scriptsize(\myhigh{-17.76})} & 67.88 {\color[rgb]{0.0,0.5,0.0}\scriptsize(-5.36)} & 71.34 {\color[rgb]{0.0,0.5,0.0}\scriptsize(-1.90)} & 75.27 {\color[rgb]{1,0,0}\scriptsize(+2.03)} & 70.89 {\color[rgb]{0.0,0.5,0.0}\scriptsize(-2.35)} \\
& Qwen2.5 14B & 78.88 & 71.45 {\color[rgb]{0.0,0.5,0.0}\scriptsize(-7.43)} & 71.08 {\color[rgb]{0.0,0.5,0.0}\scriptsize(-7.80)} & 74.11 {\color[rgb]{0.0,0.5,0.0}\scriptsize(-4.77)} & 77.95 {\color[rgb]{0.0,0.5,0.0}\scriptsize(-0.93)} & 71.26 {\color[rgb]{0.0,0.5,0.0}\scriptsize(-7.62)} \\
& Qwen2.5 7B & 67.87 & 65.60 {\color[rgb]{0.0,0.5,0.0}\scriptsize(-2.27)} & 57.47 {\color[rgb]{0.0,0.5,0.0}\scriptsize(\myhigh{-10.40})} & 58.01 {\color[rgb]{0.0,0.5,0.0}\scriptsize(-9.86)} & 62.76 {\color[rgb]{0.0,0.5,0.0}\scriptsize(-5.11)} & 57.66 {\color[rgb]{0.0,0.5,0.0}\scriptsize(\myhigh{-10.21})} \\
& StarCoder2 15B & 51.33 & 49.05 {\color[rgb]{0.0,0.5,0.0}\scriptsize(-2.28)} & 42.07 {\color[rgb]{0.0,0.5,0.0}\scriptsize(-9.26)} & 38.40 {\color[rgb]{0.0,0.5,0.0}\scriptsize(\myhigh{-12.93})} & 29.68 {\color[rgb]{0.0,0.5,0.0}\scriptsize(\myhigh{-21.65})} & 34.89 {\color[rgb]{0.0,0.5,0.0}\scriptsize(\myhigh{-16.44})} \\
& StarCoder2 7B & 54.71 & 40.08 {\color[rgb]{0.0,0.5,0.0}\scriptsize(\myhigh{-14.63})} & 45.31 {\color[rgb]{0.0,0.5,0.0}\scriptsize(-9.40)} & 33.90 {\color[rgb]{0.0,0.5,0.0}\scriptsize(\myhigh{-20.81})} & 22.32 {\color[rgb]{0.0,0.5,0.0}\scriptsize(\myhigh{-32.39})} & 34.52 {\color[rgb]{0.0,0.5,0.0}\scriptsize(\myhigh{-20.19})} \\
& StarCoder2 3B & 48.15 & 36.31 {\color[rgb]{0.0,0.5,0.0}\scriptsize(\myhigh{-11.84})} & 34.56 {\color[rgb]{0.0,0.5,0.0}\scriptsize(\myhigh{-13.59})} & 19.91 {\color[rgb]{0.0,0.5,0.0}\scriptsize(\myhigh{-28.24})} & 10.74 {\color[rgb]{0.0,0.5,0.0}\scriptsize(\myhigh{-37.41})} & 18.28 {\color[rgb]{0.0,0.5,0.0}\scriptsize(\myhigh{-29.87})} \\
\bottomrule
\end{tabular}
\end{table*}

\subsection{More Results on Multiple Mutated Experiments}
More experimental results on single mutated benchmark are displayed in \Cref{tab:ava_mul_performance}.

\begin{table*}[htbp]
\centering
\setlength{\tabcolsep}{7pt}
\caption{Model Pass@1 performance on original and multiple mutated Avatar and TransCoder benchmarks. Performance drops of more than 20\% are highlighted.}
\label{tab:ava_mul_performance}
\begin{tabular}{l|l|c|lll}
\toprule
\textbf{Type} & \textbf{Model} & \textbf{Origin} & \textbf{FUV} & \textbf{AUV} & \textbf{AFU} \\
\midrule
\multirow{10}{*}{Avatar} 
& DeepSeek V3 & 80.64 & 70.24 {\color[rgb]{0.0,0.5,0.0}\scriptsize(-10.40)} & 74.75 {\color[rgb]{0.0,0.5,0.0}\scriptsize(-5.89)} & 64.39 {\color[rgb]{0.0,0.5,0.0}\scriptsize(-16.25)} \\
& GPT-4o mini & 72.48 & 51.95 {\color[rgb]{0.0,0.5,0.0}\scriptsize(\myhigh{-20.53})} & 61.62 {\color[rgb]{0.0,0.5,0.0}\scriptsize(-10.86)} & 41.16 {\color[rgb]{0.0,0.5,0.0}\scriptsize(\myhigh{-31.32})} \\
& Llama 3.1 70B & 72.80 & 48.29 {\color[rgb]{0.0,0.5,0.0}\scriptsize(\myhigh{-24.51})} & 54.34 {\color[rgb]{0.0,0.5,0.0}\scriptsize(-18.46)} & 42.58 {\color[rgb]{0.0,0.5,0.0}\scriptsize(\myhigh{-30.22})} \\
& Llama 3.1 8B & 46.48 & 20.73 {\color[rgb]{0.0,0.5,0.0}\scriptsize(\myhigh{-25.75})} & 18.79 {\color[rgb]{0.0,0.5,0.0}\scriptsize(\myhigh{-27.69})} & 12.65 {\color[rgb]{0.0,0.5,0.0}\scriptsize(\myhigh{-33.83})} \\
& Qwen2.5-Coder 32B & 75.64 & 57.32 {\color[rgb]{0.0,0.5,0.0}\scriptsize(\myhigh{-18.32})} & 52.32 {\color[rgb]{0.0,0.5,0.0}\scriptsize(\myhigh{-23.32})} & 39.48 {\color[rgb]{0.0,0.5,0.0}\scriptsize(\myhigh{-36.16})} \\
& Qwen2.5-Coder 14B & 65.20 & 49.02 {\color[rgb]{0.0,0.5,0.0}\scriptsize(-16.18)} & 46.06 {\color[rgb]{0.0,0.5,0.0}\scriptsize(\myhigh{-19.14})} & 36.39 {\color[rgb]{0.0,0.5,0.0}\scriptsize(\myhigh{-28.81})} \\
& Qwen2.5-Coder 7B & 64.60 & 34.88 {\color[rgb]{0.0,0.5,0.0}\scriptsize(\myhigh{-29.72})} & 41.01 {\color[rgb]{0.0,0.5,0.0}\scriptsize(\myhigh{-23.59})} & 24.13 {\color[rgb]{0.0,0.5,0.0}\scriptsize(\myhigh{-40.47})} \\
& StarCoder2 15B & 58.12 & 19.76 {\color[rgb]{0.0,0.5,0.0}\scriptsize(\myhigh{-38.36})} & 25.25 {\color[rgb]{0.0,0.5,0.0}\scriptsize(\myhigh{-32.87})} & 20.16 {\color[rgb]{0.0,0.5,0.0}\scriptsize(\myhigh{-37.96})} \\
& StarCoder2 7B & 40.00 & 14.39 {\color[rgb]{0.0,0.5,0.0}\scriptsize(\myhigh{-25.61})} & 15.15 {\color[rgb]{0.0,0.5,0.0}\scriptsize(\myhigh{-24.85})} & 12.00 {\color[rgb]{0.0,0.5,0.0}\scriptsize(\myhigh{-28.00})} \\
& StarCoder2 3B & 41.80 & 10.00 {\color[rgb]{0.0,0.5,0.0}\scriptsize(\myhigh{-31.80})} & 16.57 {\color[rgb]{0.0,0.5,0.0}\scriptsize(\myhigh{-25.23})} & 8.00 {\color[rgb]{0.0,0.5,0.0}\scriptsize(\myhigh{-33.80})} \\

\midrule
\multirow{10}{*}{TransCoder} 
& DeepSeek V3 & 79.36 & 62.47 {\color[rgb]{0.0,0.5,0.0}\scriptsize(-16.89)} & 66.22 {\color[rgb]{0.0,0.5,0.0}\scriptsize(-13.79)} & 72.91 {\color[rgb]{0.0,0.5,0.0}\scriptsize(-6.45)} \\
& GPT-4o mini & 78.11 & 45.16 {\color[rgb]{0.0,0.5,0.0}\scriptsize(\myhigh{-32.95})} & 49.05 {\color[rgb]{0.0,0.5,0.0}\scriptsize(\myhigh{-29.06})} & 57.09 {\color[rgb]{0.0,0.5,0.0}\scriptsize(\myhigh{-21.44})} \\
& Llama 3.1 70B & 78.88 & 51.94 {\color[rgb]{0.0,0.5,0.0}\scriptsize(\myhigh{-26.94})} & 58.34 {\color[rgb]{0.0,0.5,0.0}\scriptsize(\myhigh{-20.54})} & 59.27 {\color[rgb]{0.0,0.5,0.0}\scriptsize(\myhigh{-24.89})} \\
& Llama 3.1 8B & 72.56 & 25.16 {\color[rgb]{0.0,0.5,0.0}\scriptsize(\myhigh{-47.40})} & 31.51 {\color[rgb]{0.0,0.5,0.0}\scriptsize(\myhigh{-41.05})} & 22.82 {\color[rgb]{0.0,0.5,0.0}\scriptsize(\myhigh{-49.74})} \\
& Qwen2.5-Coder 32B & 73.24 & 56.25 {\color[rgb]{0.0,0.5,0.0}\scriptsize(-17.00)} & 54.58 {\color[rgb]{0.0,0.5,0.0}\scriptsize(\myhigh{-25.48})} & 58.55 {\color[rgb]{0.0,0.5,0.0}\scriptsize(\myhigh{-20.04})} \\
& Qwen2.5-Coder 14B & 78.88 & 52.36 {\color[rgb]{0.0,0.5,0.0}\scriptsize(\myhigh{-26.52})} & 57.72 {\color[rgb]{0.0,0.5,0.0}\scriptsize(\myhigh{-21.16})} & 58.55 {\color[rgb]{0.0,0.5,0.0}\scriptsize(\myhigh{-20.33})} \\
& Qwen2.5-Coder 7B & 67.87 & 27.30 {\color[rgb]{0.0,0.5,0.0}\scriptsize(\myhigh{-40.57})} & 42.75 {\color[rgb]{0.0,0.5,0.0}\scriptsize(\myhigh{-25.12})} & 45.98 {\color[rgb]{0.0,0.5,0.0}\scriptsize(\myhigh{-21.89})} \\
& StarCoder2 15B & 51.33 & 12.79 {\color[rgb]{0.0,0.5,0.0}\scriptsize(\myhigh{-38.54})} & 29.97 {\color[rgb]{0.0,0.5,0.0}\scriptsize(\myhigh{-21.36})} & 23.64 {\color[rgb]{0.0,0.5,0.0}\scriptsize(\myhigh{-27.69})} \\
& StarCoder2 7B & 54.71 & 7.56  {\color[rgb]{0.0,0.5,0.0}\scriptsize(\myhigh{-47.15})} & 8.92 {\color[rgb]{0.0,0.5,0.0}\scriptsize(\myhigh{-45.79})} & 11.27 {\color[rgb]{0.0,0.5,0.0}\scriptsize(\myhigh{-43.44})} \\
& StarCoder2 3B & 48.15 & 3.32 {\color[rgb]{0.0,0.5,0.0}\scriptsize(\myhigh{-44.83})} & 9.41 {\color[rgb]{0.0,0.5,0.0}\scriptsize(\myhigh{-38.74})} & 13.84 {\color[rgb]{0.0,0.5,0.0}\scriptsize(\myhigh{-34.31})} \\

\bottomrule
\end{tabular}
\end{table*}

\section{Prompts}
In this section, we present the prompts we use to instruct the model to do code execution tasks and code translation tasks in our experiments.

\begin{center}
\fcolorbox{black}{gray!10}{\parbox{.9\linewidth}{Based on the given Python code, which may contain errors, complete the assert statement with the output when executing the code on the given test case. Do NOT output any extra information, even if the function is incorrect or incomplete. Output ``\# done'' after the assertion.
}}
\end{center}

\begin{center}
\fcolorbox{black}{gray!10}{\parbox{.9\linewidth}{You are a code translation expert. Translate the Python code below to Java. Do NOT output any extra information.
}}
\end{center}

\section{Case Study}

\begin{figure}[ht] 
    \centering
    \subfigure[CRUXEval example generated by GPT-4o mini.]{
        \includegraphics[width=0.8\textwidth]{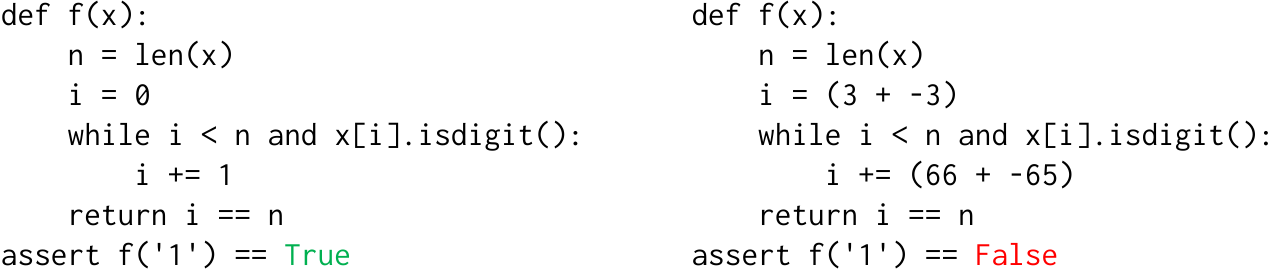}
        \label{fig:case_1_1} 
    } \\
    \subfigure[CRUXEval example generated by Qwen2.5-Coder 32B]{
        \includegraphics[width=0.8\textwidth]{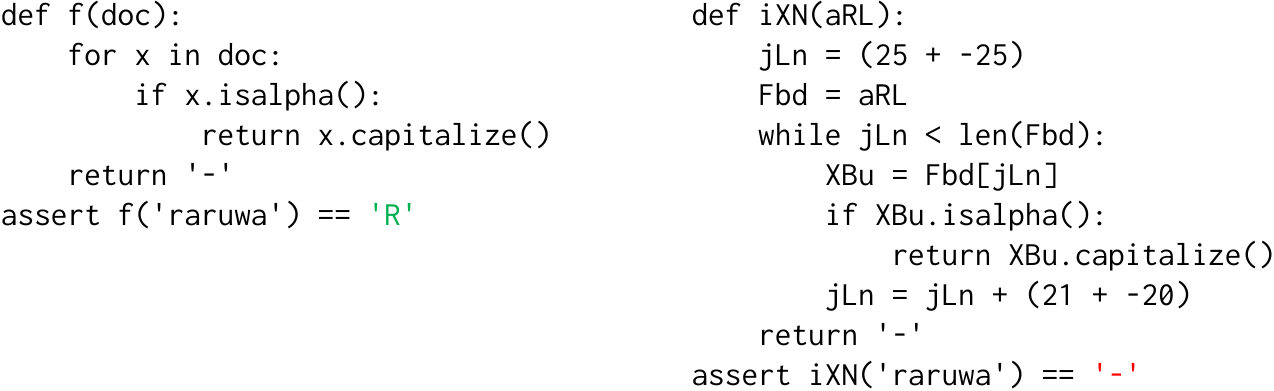}
        \label{fig:case_1_2} 
    } \\
    \caption{Case studies of code execution.} 
    \label{fig:case_1} 
\end{figure}

\begin{figure}[ht] 
    \centering
    \subfigure[Avatar example generated by DeepSeek V3.]{
        \includegraphics[width=\textwidth]{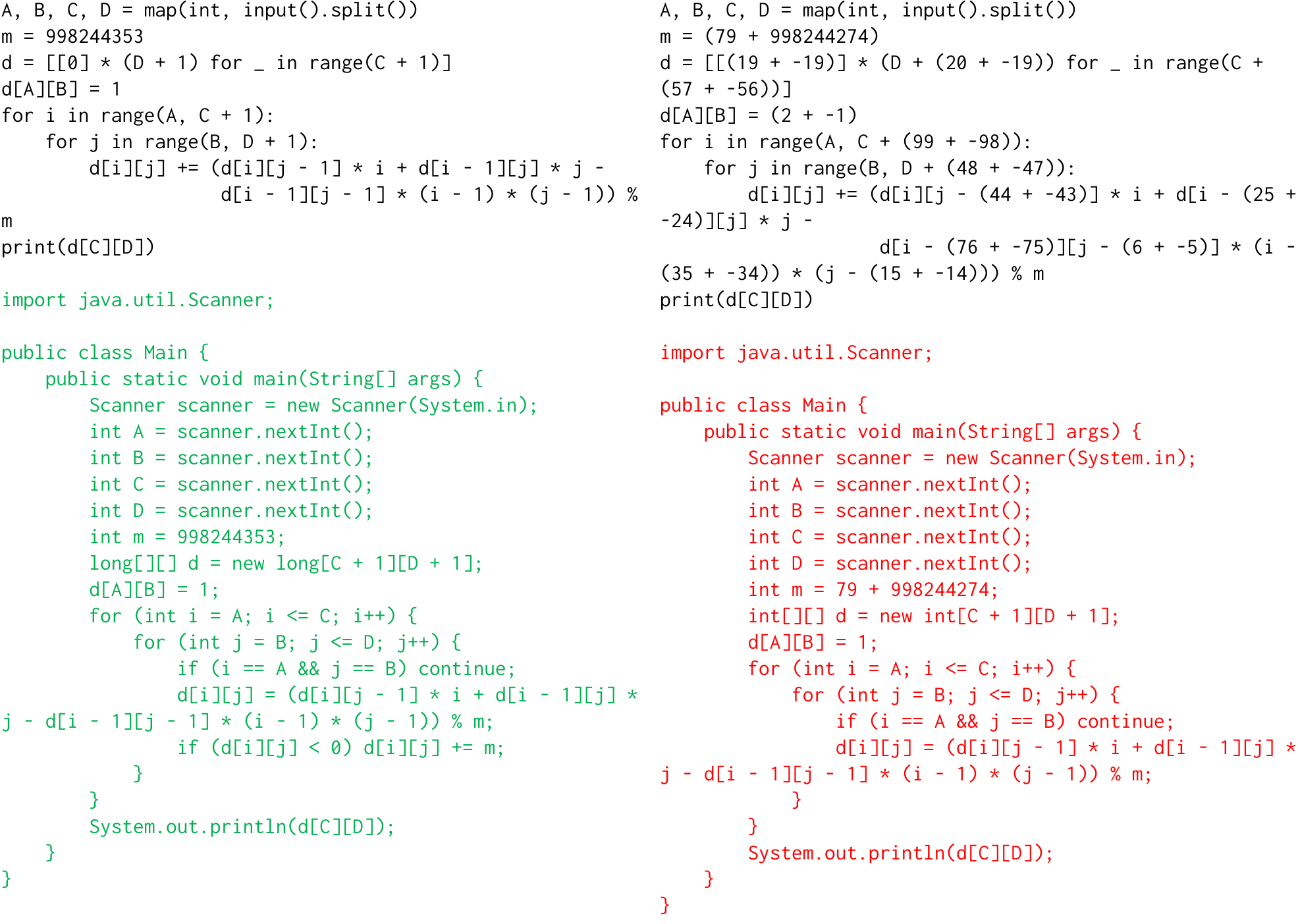}
        \label{fig:case_2_1} 
    } \\
    \subfigure[TransCoder example generated by Llama 3.1 8B.]{
        \includegraphics[width=\textwidth]{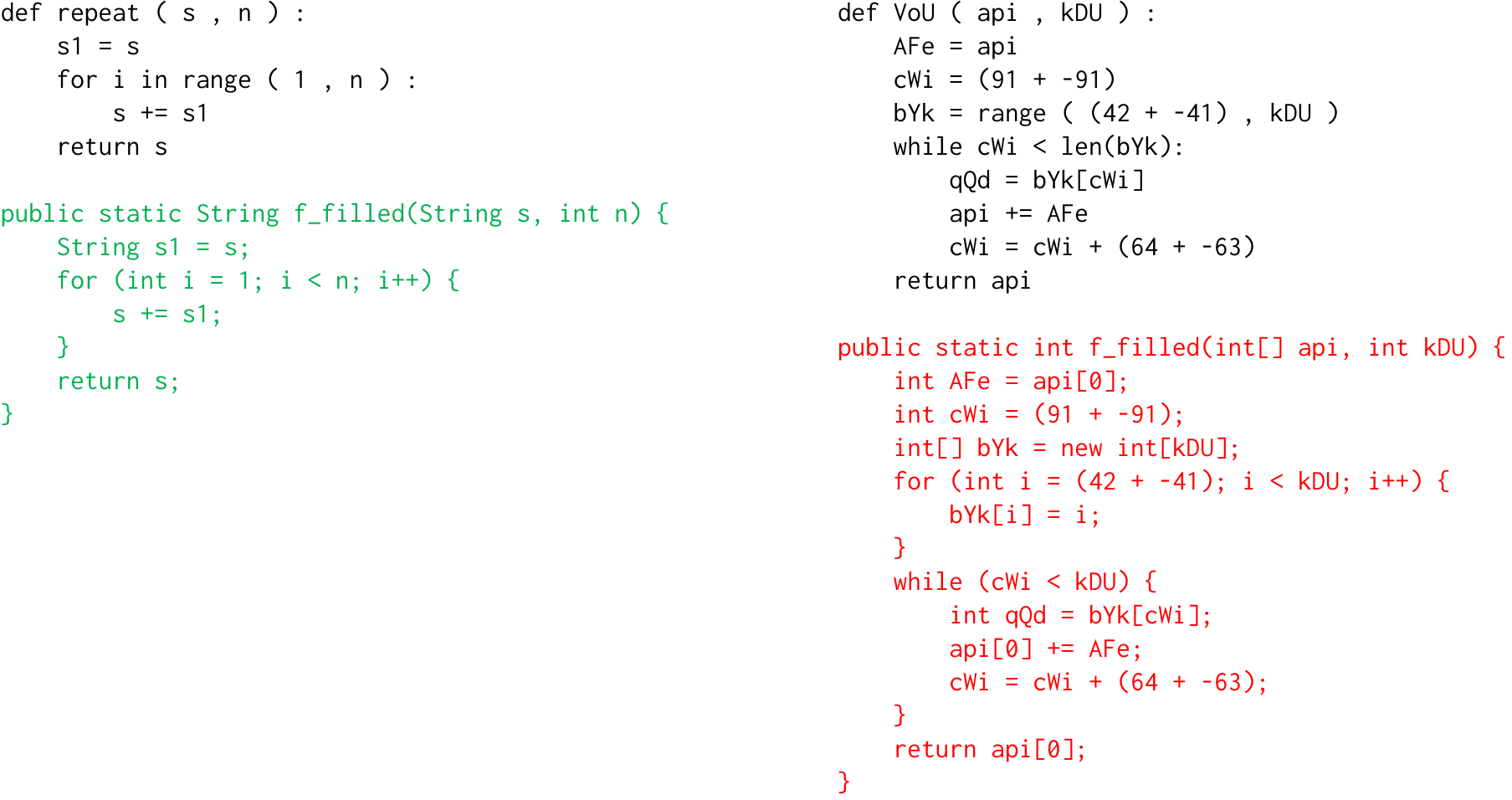}
        \label{fig:case_2_2} 
    } \\
    \caption{Case studies of code translation.} 
    \label{fig:case_2} 
\end{figure}

In this section, we present some cases generated in our experiments in \Cref{fig:case_1} and \Cref{fig:case_2}.

\Cref{fig:case_1} shows two examples of code execution tasks performed by the GPT-4o mini and Qwen2.5-Coder 32B models. In \Cref{fig:case_1_1}, the LLM was originally able to make the correct judgment, but after applying the Constant Unfolding mutation, it made an incorrect judgment. \Cref{fig:case_1_2} presents a slightly more complex example. After the FUV mutation, the model incorrectly evaluated the \texttt{isalpha} condition and output an erroneous result.

\Cref{fig:case_2} shows two examples of code translation tasks performed by the DeepSeek V3 and Llama 3.1 8B models. In \Cref{fig:case_2_1}, the key to correct translation lies in paying attention to the sign of the numbers in the array. Since Python and Java define the modulo operation for negative numbers differently, this detail must be carefully considered when translating from Python to Java. In the original code, the LLM successfully recognized this aspect. However, after applying the Constant Unfolding mutation, the LLM overlooked this detail and ultimately produced code that could not pass the test. Additionally, since Python does not explicitly specify types, correctly determining the type is also a crucial aspect of translating from Python to Java. In \Cref{fig:case_2_2}, the original code successfully identifies the most suitable type for the code. However, after the FUV mutation, the model makes an incorrect judgment.

\end{document}